\RequirePackage[2020-02-02]{latexrelease}
\documentclass[amssymb,prb,twocolumn,showpacs]{revtex4}
\usepackage{epsfig}
\usepackage{dcolumn}
\usepackage{amsmath}
\begin{document}
\title{\bf\ Coherent states in microwave-induced resistance oscillations and zero resistance states.}
\author{J. I\~narrea$^{1,3}$ and G. Platero$^{2,3}$}

\address{$^1$Escuela Polit\'ecnica
Superior,Universidad Carlos III,Leganes, Madrid, 28911, Spain\\
$^2$Instituto de Ciencia de Materiales, CSIC,
Cantoblanco, Madrid, 28049, Spain.\\
$^3$Unidad Asociada al Instituto de Ciencia de Materiales, CSIC,
Cantoblanco, Madrid, 28049, Spain.}

\begin{abstract}
We investigate irradiated high-mobility two-dimensional electron systems
(2DES) under low or moderated magnetic fields.
These systems present microwave-induced magnetoresistance oscillations (MIRO) which,  as we demonstrate,
 reveal the presence of coherent states of the quantum harmonic oscillator.
We also show that the principle of
minimum uncertainty of coherent states is at the heart of
MIRO and zero resistance states (ZRS).
Accordingly, we are able to explain, based on coherent states, important experimental evidence of these
  photo-oscillations.
Such as their  physical origin, their periodicity with the
inverse of the magnetic field and  their peculiar oscillations minima and maxima positions
in regards of the magnetic field.
Thus, remarkably enough, we come to the conclusion that 2DES, under low  magnetic fields, become a system of quasiclassical states or
coherent states and MIRO would be the smoking gun of the existence of these peculiar states is these systems.

\end{abstract}
\maketitle
{\it Introduction.} The first idea of coherent states or quasiclassical states was introduced by Schr\"odinger\cite{sro} describing
minimum uncertainty constant-shape Gaussian wave packets of the quantum harmonic oscillator.
They were constructed by the quantum superposition of the stationary states of the harmonic oscillator. These
wave packets
\begin{figure}
\centering\epsfxsize=2.5in \epsfysize=2.5in
\epsffile{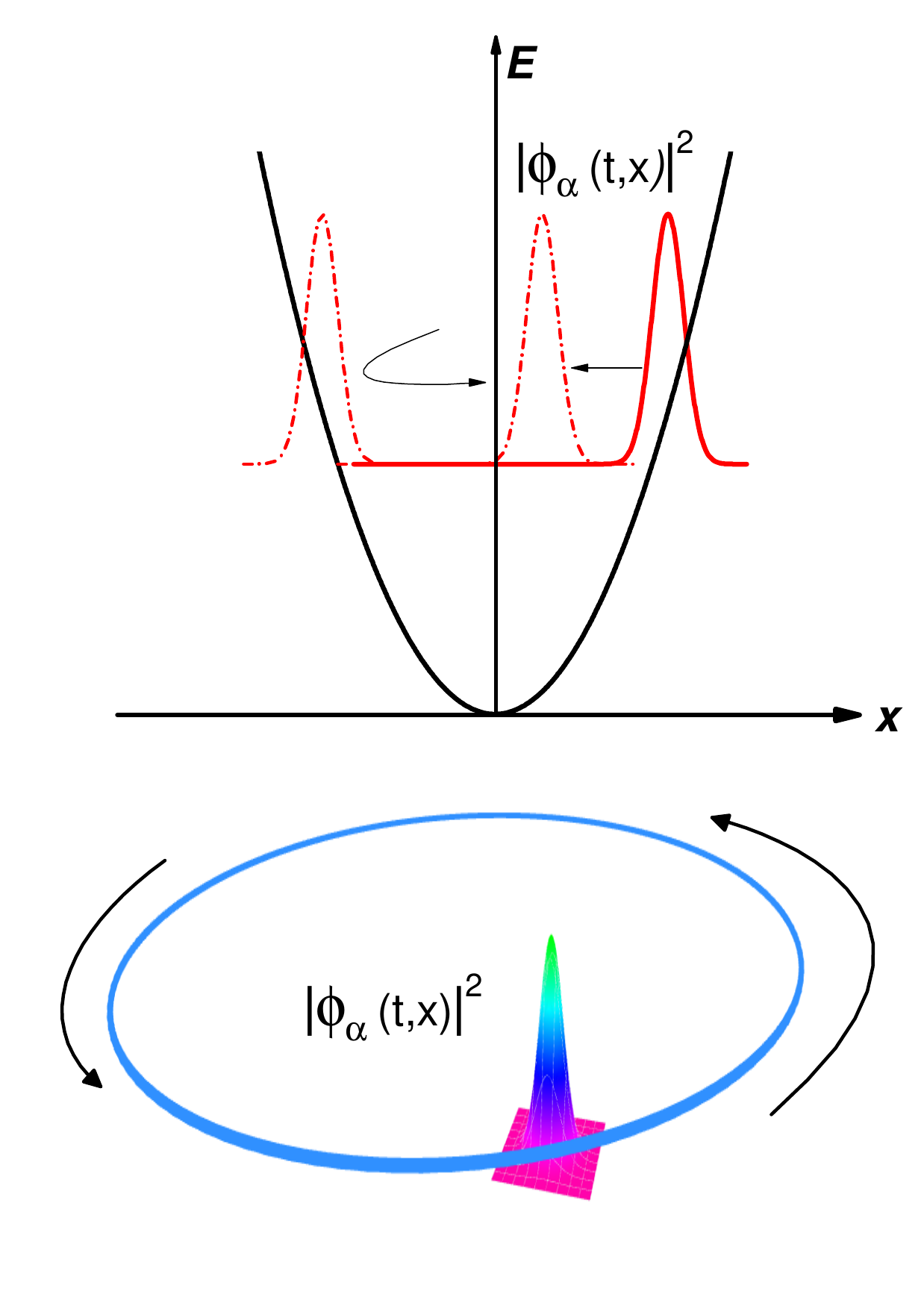}
\caption{Schematic diagrams of coherent states: The probability density of the coherent state is a
constant-shaped Gaussian
distribution, whose center oscillates in a harmonic potential similarly as
its classical counterpart. The lower part exhibits the 2D approach.}
\end{figure}
displaced harmonically oscillating similarly as their classical counterpart\cite{sro}.
Later on, Glauber\cite{glau} applied the concept of coherent states to the
electromagnetic field being described by a sum of quantum field oscillators for
each field frequency or mode. These coherent states of electromagnetic radiation introduced by Glauber are
extensively used nowadays in quantum optics.
Coherent states\cite{dodonov,yurke,noel,dodonov2} are also an essential and powerful  tool in condensed
matter when describing the dynamics of
quantum systems that are very close to a classical behaviour.
One remarkable example of this consists of one electron under
the influence of a moderate and constant magnetic field ($B$).
The quantum mechanical solution of this problem leads us to
Landau states which are mere stationary states of the quantum harmonic oscillator.
Under low or moderate values of $B$, this system can be described by
an infinite superposition of Landau states, i.e., a coherent state.
The resulting wave packet oscillates classically at the cyclotron frequency ($w_{c}$)
inside the quadratic potential keeping constant the Gaussian shape (see Fig. 1)
and complying with the minimum uncertainty condition.

The discovery of MIRO two decades ago  led to a great deal of
theoretical works back then as
the displacement model
\cite{girvin}, the inelastic model\cite{dimi}
and the microwave-driven
electron orbits model\cite{ina1,ina2,ina3,ina4}. According to the latter,
Landau states, under radiation, spatially and  harmonically  oscillate with the guiding center at the radiation frequency ($w$) performing
 classical trajectories. In this swinging motion electrons
 are  scattered by charged impurities giving rise to oscillations in the irradiated magnetoresistance, i.e., MIRO.

In this letter we demonstrate that the electron dynamics  and magnetotransport in
high-mobility 2DES is governed by the coherent
states of the
quantum harmonic oscillator.
In fact, we conclude that 2DES under low or moderate $B$ become a systems of
coherent states and when irradiated, MIRO\cite{mani,zudov,ryzhii} bring to light the peculiar nature of these states.
In other words,  irradiated coherent states of the quantum harmonic oscillator are
at the heart of  MIRO.
Accordingly, we incorporate the concept of coherent states to the microwave-driven electron orbit model\cite{ina1,ina2,ina3,ina4}.
Thus, a remarkable obtained result  is  that the time, $\tau$ (evolution time\cite{cohen}),
it takes a scattered electron to jump between coherent states
to give significant contributions to the current
has to be equal to
the cyclotron period, $T_{c}=2\pi/w_{c}$.
For different values of $\tau$, the contribution turns out negligible.
This result holds in the dark
and under radiation where $\tau$ will play an essential role. Thus,  MIRO
is mainly dependent on $\tau$ along with $w$.
$\tau$  also determines the peculiar $B$-dependent MIRO extrema position
and explain the periodicity of MIRO with the inverse of $B$.
Thus, MIRO   finally reveals that
coherent states of the quantum harmonic oscillator are present in
high-mobility 2DES when under low $B$ playing a lead role in magnetotransport both in the dark and
under radiation.
On the other hand, coherent states minimize the Heisenberg uncertainty principle and then, in our model
this would establish which
states can be reached by scattering.

\begin{figure}
\centering\epsfxsize=2.8in \epsfysize=2.8in
\epsffile{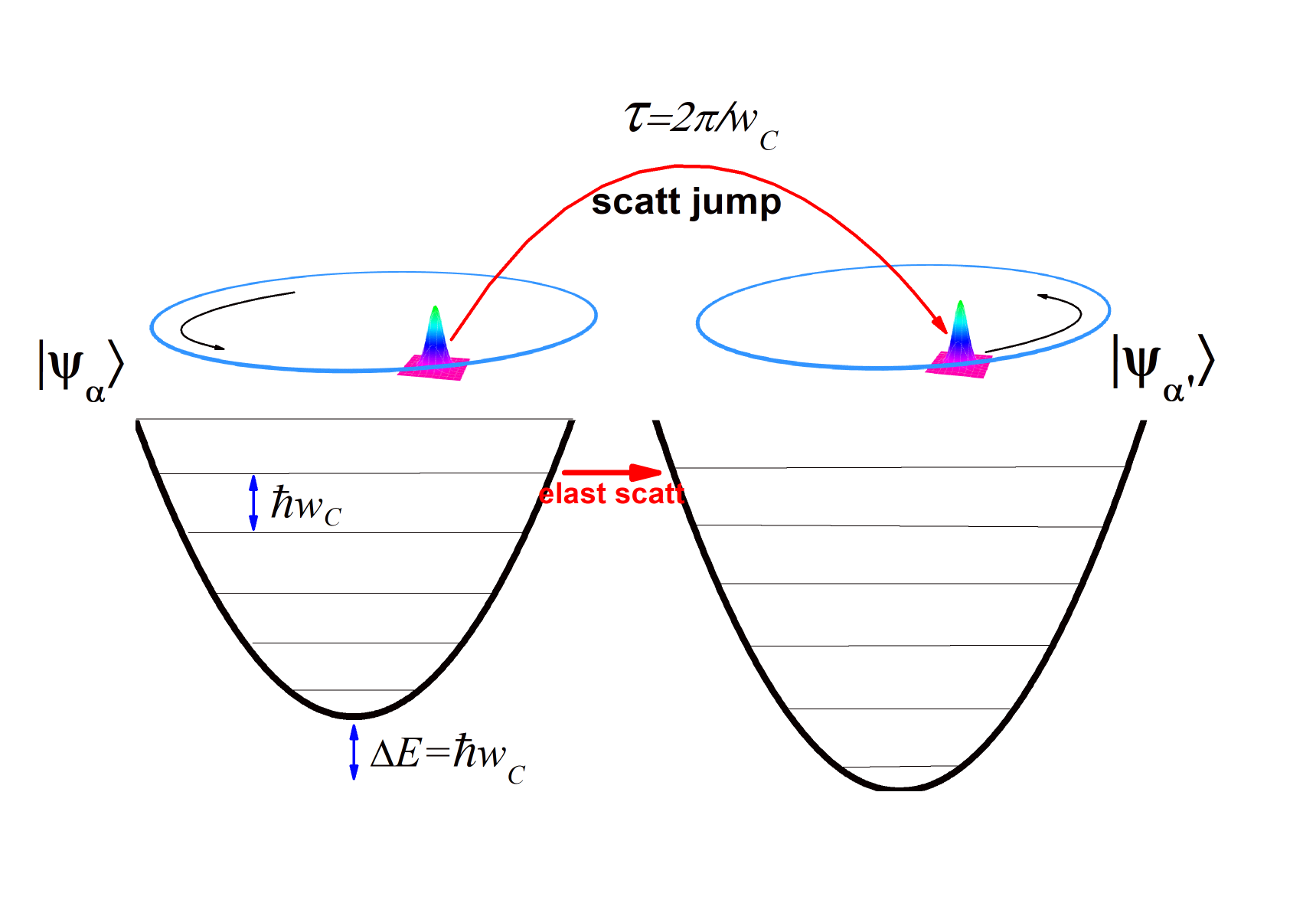}
\caption{Schematic diagram of scattering process between  coherent states $\Psi_{\alpha}$
and $\Psi_{\alpha'}$. The scattering is quasi-elastic.  The probability density for both
 coherent states is a
constant-shaped Gaussian wave packet. The process
evolution time, $\tau$ is the cyclotron period. i.e., $\tau=2\pi /w_{c}=T_{c}$.
$\Delta E$ is the energy difference between the coherent states.
}
\end{figure}
{\it Theoretical model.}
We first obtain
an expression for the coherent states of a radiation-driven
quantum harmonic oscillator. The starting point is the exact solution
of the time-dependent Schr\"odinger equation of a quantum harmonic oscillator under
a time-dependent force. This corresponds to
the electronic wave function for a 2DES in a
perpendicular $B$, a DC electric field, $E_{DC}$, and MW radiation which is
considered semi-classically.
 The total hamiltonian $H$ can be written as:
\begin{eqnarray}
H&=&\frac{P_{x}^{2}}{2m^{*}}+\frac{1}{2}m^{*}w_{c}^{2}(x-X(0))^{2}-eE_{dc}X(0) +\nonumber \\
 & &+\frac{1}{2}m^{*}\frac{E_{dc}^{2}}{B^{2}}\nonumber-eE_{0}\cos wt (x-X(0)) -\nonumber \\
 & &-eE_{0}\cos wt X(0) \nonumber\\
 &=&H_{1}-eE_{0}\cos wt X(0)
\end{eqnarray}
 where the corresponding wave function solution is given by\cite{ina1,kerner,park}:
\begin{equation}
\Psi_{n}(x,t)=\phi_{n}(x-X(0)-x_{o}(t)) e^{-i w_{c} (n+1/2)t} e^{\frac{i}{\hbar}\Theta (t)}
\end{equation}
where,
\begin{eqnarray}
\Theta (t)&=&\left[m^{*}\frac{dx_{o}(t)}{dt}x-
\int_{0}^{t} {\it L} dt' \right]\nonumber\\ +&& X(0)\left[-m^{*}\frac{dx_{o}(t)}{dt}x+
\int_{0}^{t} {\it E_{0} \cos wt'} dt' \right]
\end{eqnarray}
 $X(0)$ is the guiding center of the driven-Landau state,
$E_{0}$ the MW electric field intensity,
 $\phi_{n}$
is the solution for the Schr\"{o}dinger equation of the unforced
quantum harmonic oscillator  and $x_{0}(t)$ is the classical
solution of a forced harmonic oscillator:
\begin{equation}
x_{0}(t)=\frac{e E_{o}}{m^{*}\sqrt{(w_{c}^{2}-w^{2})^{2}+w^{2}\gamma^{2}}}\sin wt
=A\sin wt
\end{equation}
where $\gamma$ is a phenomenologically-introduced damping factor
for the electronic interaction with acoustic phonons and
${L}$ is the classical Lagrangian.
Apart from phase factors, the wave function turns out to be
the same as a quantum harmonic oscillator (Landau state) where the center is
driven by $x_{0}(t)$. Thus, all driven-Landau states harmonically oscillates   in phase
at the radiation frequency.

A coherent state denoted by $| \alpha \rangle$ is defined as the eigenvector of the annihilation operator $\hat{a}$
with eigenvalue $\alpha$
and can  be expressed as a superposition of quantum harmonic oscillator states\cite{cohen},
\begin{equation}
| \alpha \rangle=\sum_{n}c_{n}(\alpha)|\phi_{n}\rangle=e^{-|\alpha|^{2}/2}\sum_{n}\frac{\alpha^{n}}{\sqrt{n!}}|\phi_{n}\rangle
\end{equation}
The coherent state  $| \alpha \rangle$ can be also obtained  with the displacement operator $D(\alpha)$\cite{cohen}
acting on the quantum harmonic oscillator  ground state $|\phi_{0}\rangle$, $| \alpha \rangle= D(\alpha)|\phi_{0}\rangle$,
where the unitary operator $D(\alpha)$ is defined by: $D(\alpha)=e^{\alpha a^{\dagger}-\alpha^{\ast}a}$.
The coherent state in the position representation or wave function then reads,
$\psi_{\alpha}(x)=\langle x| D(\alpha)|\phi_{0}\rangle$.
We observe, according to the obtained
MW-driven wave function  (Eq. 2),  that the irradiated Landau level structure remains unchanged with
respect to the undriven situation; same Landau level index and energy. Then,
we conclude that the system is quantized, in the same way as the
unforced quantum harmonic oscillator\cite{park}.
 Thus, we can construct the driven-coherent
states based on  driven-Landau states similarly as if they were undriven\cite{cohen}:
\begin{eqnarray}
|\psi_{\alpha}(x,t) \rangle&=&e^{\frac{i}{\hbar}\Theta (t)} e^{-iw_{c}t/2} e^{-|\alpha|^{2}/2}\nonumber\\
&\times&\sum_{n}\frac{(\alpha e^{-iw_{c}t})^{n} }{\sqrt{n!}}|\phi_{n}(x-X(0)-x_{o}(t)       )\rangle\nonumber\\
\end{eqnarray}
Now applying the displacement operator, we can calculate the wave function corresponding to
the coherent state of the MW-driven quantum oscillator:

\begin{eqnarray}
\psi_{\alpha}(x,t)=e^{\frac{i}{\hbar}\Theta (t)}e^{-iw_{c}t/2}\langle x| D(\alpha)|\phi_{0}(x-X(0)-x_{o}(t))\rangle&&\nonumber\\
\nonumber\\
= e^{\frac{i}{\hbar}\Theta (t)} e^{i\vartheta_{\alpha}}e^{-iw_{c}t/2}e^{\frac{i}{\hbar}\langle p \rangle(t)x}
\phi_{0}[x-X(0)-x_{o}(t)-\langle x \rangle(t)]\nonumber\\
\end{eqnarray}
where,
\begin{equation}
\phi_{0}[x-X(0)-x_{o}(t)-\langle x \rangle(t)]=
\left (\frac{mw_{c}}{\pi\hbar}\right )^{1/4}
e^{-\left[\frac{x-X(0)-x_{o}(t)-\langle x \rangle(t)}{2\Delta x}\right]^{2}}
\end{equation}
$\langle x \rangle(t)$ and $\langle p  \rangle(t)$ are the position and
momentum mean values respectively\cite{cohen},
$\langle x \rangle(t)=\sqrt{\frac{2\hbar}{m^{\ast} w_{c}}}|\alpha_{0}|\cos(w_{c}t-\varphi)$
 and
$\langle p  \rangle(t)=-\sqrt{2m^{\ast}\hbar w_{c}}|\alpha_{0}|\sin(w_{c}t-\varphi)$
where we  have used that  $\alpha=|\alpha_{0}| e^{-(iw_{c}t-\varphi)}$.
$\Delta x$
is
the position uncertainty and the global phase factor, $ e^{i\vartheta_{\alpha}}=e^{\alpha^{\ast 2}-\alpha^{2}}$.
Then, the wave packet associated with $\Psi_{\alpha}(x,t)$ is therefore given by:
\begin{equation}
|\Psi_{\alpha}(x,t)|^{2}=|\phi_{0}[x-X(0)-x_{o}(t)-\langle x \rangle(t)]|^{2}
\end{equation}
Thus, according to the above,
the microscopic physical description of a high-mobility 2DES under low or moderate $B$ would consist of
constant-shaped Gaussian wave packets harmonically displacing with $w_{c}$ in
the undriven case and with $w_{c}$ and  $w$ under radiation.

To calculate the longitudinal magnetoresistance, $R_{xx}$, we first obtain the longitudinal conductivity ${\sigma_{xx}}$ following
a semiclassical Boltzmann model\cite{ridley,ando,askerov},
 \begin{equation}
\sigma_{xx}=2e^{2} \int_{0}^{\infty} dE \rho_{i}(E) (\Delta X_{0})^{2}W_{I}\left( -\frac{df(E)}{dE}  \right)
\end{equation}
being $E$ the energy, $\rho_{i}(E)$ the Landau states density of the
initial coherent state and $W_{I}$ is the electron-charged impurities scattering rate.
We consider now that the scattering takes place between coherent states of quantum harmonic oscillators.
Thus, $\Delta X_{0}$ is the distance between the guiding centers  of the scattering-involved coherent states.

We first study  the dark case and
according to the Fermi's golden rule $W_{I}$ is given by,
\begin{equation}
  W_{I}=N_{i}\frac{2\pi}{\hbar}|<\psi_{\alpha^{'}}|V_{s}|\psi_{\alpha}>|^{2}\delta(E_{\alpha^{'}}-E_{\alpha})
\end{equation}
where $N_{i}$ is the number of charged impurities, $\psi_{\alpha}$ and $\psi_{\alpha^{'}}$ are the wave functions  corresponding to the initial and final coherent states respectively,
 $V_{s}$ is the scattering potential for charged impurities\cite{ando}:
 $V_{s}=\sum_{q} V_{q}e^{i q_{x} x}= \sum_{q}\frac{e^{2}}{2 S \epsilon (q+q_{TF})} e^{i q_{x} x}$,
$S$ being the sample surface, $\epsilon$ the dielectric
constant, $q_{TF}$ is the Thomas-Fermi screening
constant\cite{ando} and $q_{x}$ the $x$-component of $\overrightarrow{q}$,
the electron momentum change after the  scattering event.
 $E_{\alpha}$ and $E_{\alpha^{'}}$ stand for the coherent states initial and final energies respectively.
 
The averaging on the impurities distribution has been considered in a very simple approach following Askerov\cite{askerov}, Ando\cite{ando} et. al., and J. H.Davies\cite{davies}. Thus, if the concentration of impurities is not too high, and they are randomly distributed in the sample, the interferences caused by the impurity centers can be neglected. Then, we have ignored those interreferences and assume that the scattering due to each impurity is independent of the others. As a result the total scattering is equal to the scattering rate for one impurity center multiplied by the total number of impurities $N_{i}$.

The $V_{s}$ matrix element is given by\cite{ridley,ando,askerov}:
\begin{equation}
|<\psi_{\alpha^{'}}|V_{s}|\psi_{\alpha}>|^{2}=\sum_{q}|V_{q}|^{2}|I_{\alpha,\alpha^{'}}|^{2}
\end{equation}
and the term $I_{\alpha,\alpha^{'}}$\cite{ridley,ando,askerov},
\begin{equation}
I_{\alpha,\alpha^{'}}=\int^{\infty}_{-\infty} \psi_{\alpha^{'}}(x-X^{'}(0)-\langle x' \rangle(t')) e^{i q_{x} x}\psi_{\alpha}(x-X(0)-\langle x \rangle(t)) dx
\end{equation}
After lengthy algebra we obtain an expression for $I_{\alpha,\alpha^{'}}$,
\begin{equation}
  |I_{\alpha,\alpha^{'}}|=e^{-\frac{[X^{'}(0)-X(0)+\langle x^{'} \rangle(t')-\langle x \rangle(t)]^{2}}{8(\Delta x)^{2}}}
  e^{-\frac{q_{x}^{2}(t)2(\Delta x)^{2}}{4}}
\end{equation}
where $q_{x}(t)$is given by,
\begin{eqnarray}
q_{x}(t)&=&q_{x}+\sqrt{2m\hbar w_{c}}/\hbar\left[|\alpha_{0}^{'}|\sin(w_{c}t') -|\alpha_{0}|\sin(w_{c}t)\right]\nonumber\\
&=&q_{x}+2\sqrt{2m\hbar w_{c}}/\hbar|\alpha_{0}| \cos(w_{c}(t+\tau/2))\sin(w_{c}\tau/2))\nonumber\\
\end{eqnarray}

On the other hand,
\begin{equation}
\langle x^{'} \rangle(t^{'})-\langle x \rangle(t)
\simeq \sqrt{\frac{2\hbar}{m w_{c}}}|\alpha_{0}|2\sin(w_{c}(t+\frac{\tau}{2})-\varphi) \sin(-w_{c}\frac{\tau}{2})
\end{equation}
where $t$ and  $t^{'}$ are the initial and final times for the scattering event and $\tau$ is the evolution
time between coherent states. Thus, $t^{'}=t+\tau$. We have considered also that for
low values of $B$,   $|\alpha^{'}_{0}|\simeq|\alpha_{0}|$. Developing
the above exponential we can finally get to,
\begin{equation}
  |I_{\alpha,\alpha^{'}}|\propto e^{-2|\alpha_{0}|^{2}\sin^{2}(w_{c}(t+\frac{\tau}{2})-\varphi) \sin^{2}(w_{c}\frac{\tau}{2})}
\end{equation}
For typical experimental values of $B$,  $|\alpha_{0}|^{2} > 50$ and thus, $I_{\alpha,\alpha^{'}} \rightarrow 0$.
Accordingly, the scattering rate and conductivity would be negligible  too. Nonetheless, there is an important exception when
$\tau$ equals the cyclotron period $T_{c}$:
$\tau=\frac{2\pi}{w_{c}}$.
\begin{figure}
\centering \epsfxsize=3.8in \epsfysize=4.0in
\epsffile{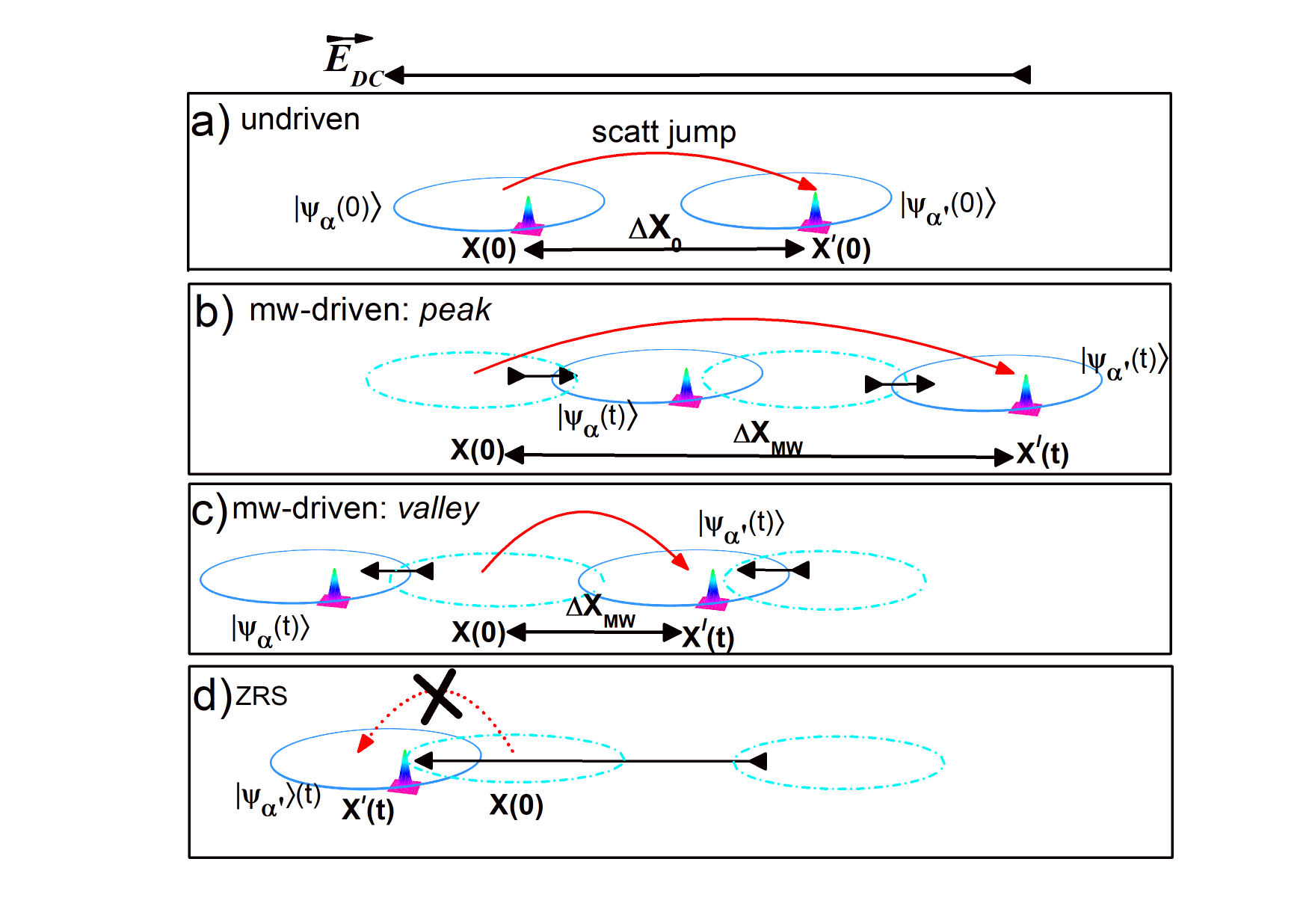}
\caption{Schematic diagrams for electron scattering between  coherent states in the dark (undriven)
and with  radiation (mw-driven).
a) Undriven  scattering. The average distance (advanced distance)  between initial $|\psi_{\alpha}\rangle$ and
final coherent state, $|\psi_{\alpha^{'}}\rangle$, is $\Delta X(0)$. This distance mainly determines $R_{xx}$.
b) MW-driven scattering giving rise to  peaks. Now the average advanced distance is larger because the final state,
minimizing the Heisenberg uncertainty principle,
is farther than the dark position due to the swinging motion of the driven-coherent states.
c)  MW-driven scattering giving rise to valleys.  When the final coherent state is  closer we obtain MIRO valleys.
 b)  Situation when MW power is high enough and the states go backwards.
In this scenario the final state ends up behind the initial state dark position and
the scattering jump can not take place. }
\end{figure}
 In other words, the scattered electron
begins and ends in the same position in the Landau orbit. Only  in this case
$I_{\alpha,\alpha^{'}} \neq 0$.
 Thus, only scattering processes fulfilling  the previous condition
of $\tau$ will efficiently contribute to the current. The rest contributions can be neglected.
Finally the expression of  $I_{\alpha,\alpha^{'}}$ reads\cite{ridley},
\begin{eqnarray}
|I_{\alpha,\alpha^{'}}|&=& e^{-\left(\frac{\left(X^{'}(0)-X(0)\right)^{2}}{8(\Delta x)^{2}}+\frac{q_{x}^{2}(\Delta x)^{2} }{2}\right)}\nonumber\\
&=&e^{-\left(\frac{q^{2}2(\Delta x)^{2}}{4}\right)}
\end{eqnarray}
where  $X^{'}(0)-X(0)=[-q_{y}2(\Delta x)^{2}]$,$\cite{ridley}$.
This, in turn, leads us to a final expression for  $W_{I}$,
\begin{equation}
W_{I}=\frac{n_{i}e^{4}}{2 \pi \hbar \epsilon^{2}}\int \frac{e^{-q^{2}(\Delta x)^{2}}}{(q+q_{TF})^{2}}(1-\cos \theta)\delta(E_{\alpha^{'}}-E_{\alpha})d^{2}q
\end{equation}
where $n_{i}$ is the  charged impurity density
and $\theta$ is the scattering
angle.
The density of initial Landau states $\rho_{i}(E)$ can be obtained by
using the Poisson sum rules to get to\cite{ihn},
$\rho_{i}(E)= \frac{m^{\ast}}{\pi \hbar^{2}}\left[1-2\cos \left(\frac{2\pi E}{\hbar w_{c}} e^{-\pi\Gamma/\hbar w_{c}} \right)\right]$.
Finally, gathering all terms and solving the energy integral,  we obtain an expression for $\sigma_{xx}$ that reads,
\begin{eqnarray}
\sigma_{xx}=\frac{n_{i}e^{6}m^{\ast}}{2\pi^{3}\hbar^{3}\epsilon^{2}}(\Delta X_{0})^{2}\frac{1}{\hbar w_{c}}\left(\frac{1+e^{-\pi\Gamma/\hbar w_{c}}}{1-e^{-\pi\Gamma/\hbar w_{c}}}\right) \times \nonumber\\
\left(1-\frac{2\chi_{s}}{\sinh (\chi_{s}) } \cos\left(\frac{2\pi E_{F}}{\hbar w_{c}}\right) e^{-\pi\Gamma/\hbar w_{c}} \right)\times\nonumber\\
 \int \frac{e^{-q^{2}(\Delta x)^{2}}}{(q+q_{TF})^{2}}(1-\cos \theta) d^{2}q
\end{eqnarray}
where $\chi_{s}=2\pi^{2}k_{B}T/\hbar w_{c}$, $k_{B}$ being the Boltzmann constant, $E_{F}$ the Fermi energy
and $\Gamma$ the Landau level width.
To obtain
$R_{xx}$ we use the relation
$R_{xx}=\frac{\sigma_{xx}}{\sigma_{xx}^{2}+\sigma_{xy}^{2}}
\simeq\frac{\sigma_{xx}}{\sigma_{xy}^{2}}$, where
$\sigma_{xy}\simeq\frac{n_{e}e}{B}$ and
$\sigma_{xx}\ll\sigma_{xy}$, $n_{e}$ being the 2D electron density.

One important condition that features coherent states is that they minimize the Heisenberg uncertainty principle. Thus, for
the time-energy uncertainty relation\cite{cohen}, $\Delta t \Delta E=h$.
For our specific problem,  $\Delta t=\tau$ that implies  $\Delta E=\hbar w_{c}$, $\Delta E$ being
 the energy difference between scattering-involved coherent states. Thus, we obtain two conditions
for the scattering between coherent states to take place, first $\tau=\frac{2\pi}{w_{c}}$ and second,
the energy difference equals  $\hbar w_{c}$.
There are also physical reasons that endorse  the latter specially in high-mobility
samples where the levels are very narrow in terms of states density.
 In these systems the only efficient contributions to scattering are the ones corresponding
 to aligned Landau levels (see Fig. 2),
i.e.,
when $\Delta E=n \times \hbar w_{c}$. The most intense  of them is when $n=1$
that corresponds to the closest in distance coherent states or smallest value
of  $\Delta X_{0}$, (see Eq. 18).
This agrees with  the condition
that when $n=1$, the Heisenberg uncertainty principle is
minimized.
The two conditions discussed above hold in the dark and under radiation.
For the latter case,  MIRO reveals, the
important role played by $\tau$ in the based-on-coherent states magnetotransport processes.

When we turn on the light, the term that is going to be mainly
affected in the $\sigma_{xx}$ expression is the distance between the coherent states
guiding
centers, i.e., $\Delta X_{0}$. This average distance now
turns into $\Delta X_{MW}$\cite{inarashba,inahole}
\begin{eqnarray}
\Delta X_{MW}&=&X^{'}_{MW}-X_{MW}\nonumber\\
&=&\Delta X_{0}-A\left(\sin w(t+\tau)-\sin wt\right)+\nonumber\\
&&\sqrt{\frac{2\hbar}{m^{\ast}w_{c}}}|\alpha_{0}|\left(\cos w_{c}(t+\tau)-\cos w_{c}t\right)\nonumber\\
\end{eqnarray}
If we consider, on average, that the scattering jump begins when the MW-driven
oscillations is at its mid-point, ($wt=2\pi n$, $n$ being a positive integer), and being $\tau=2\pi/w_{c}$,
we end up having,
\begin{equation}
\Delta X_{MW}=\Delta X_{0}-A\sin 2\pi \frac{w}{w_{c}}
\end{equation}

This result affects dramatically $\sigma_{xx}$ and in turn $R_{xx}$.
Now photo-oscillations rise according to $\Delta X_{MW}$ and its built-in sine function.
In Fig. 3 we present schematic diagrams for the different
situations regarding MIRO peaks and valleys and ZRS.
In the undriven scenario  an electron in the initial coherent state, scatters with
charged impurities  and
jumps to the final coherent state minimizing the Heisenberg uncertainty principle.
The latter condition determines what coherent states can be connected
via scattering.
On average the advanced distance is $\Delta X_{0}=X^{'}_{0}-X_{0}$,  (see Fig. 3a).
When the light is on, depending on
the term $A\sin 2\pi \frac{w}{w_{c}}$, some times the minimum uncertainty
final state will be further away  than in the dark regarding the initial state position.  Thus,
 on average,  $\Delta X_{MW}>\Delta X_{0}$ and $R_{xx}$ will be larger, giving
rise to peaks  (see Fig. 3b). On the other hand, other times the final coherent state will be closer
 and $\Delta X_{MW}<\Delta X_{0}$ and $R_{xx}$ will be smaller, giving rise to valleys, (see Fig. 3c).
  Finally, when the driven
 coherent states are going backward and the radiation power is
large enough, the final state, minimizing the uncertainty principle,
 will be behind the initial state in the dark (see Fig. 3d).
However, the scattered electron can only effectively jump forward due to the
DC electric field direction and the final coherent state can never be reached; in
the forward direction there is no final coherent state fulfilling the
 minimum uncertainty condition and the scattering can not be completed. Thus,
the system reaches the ZRS scenario where the electron remains in the initial
coherent state.

\begin{figure}
\centering \epsfxsize=3.8in \epsfysize=3.6in
\epsffile{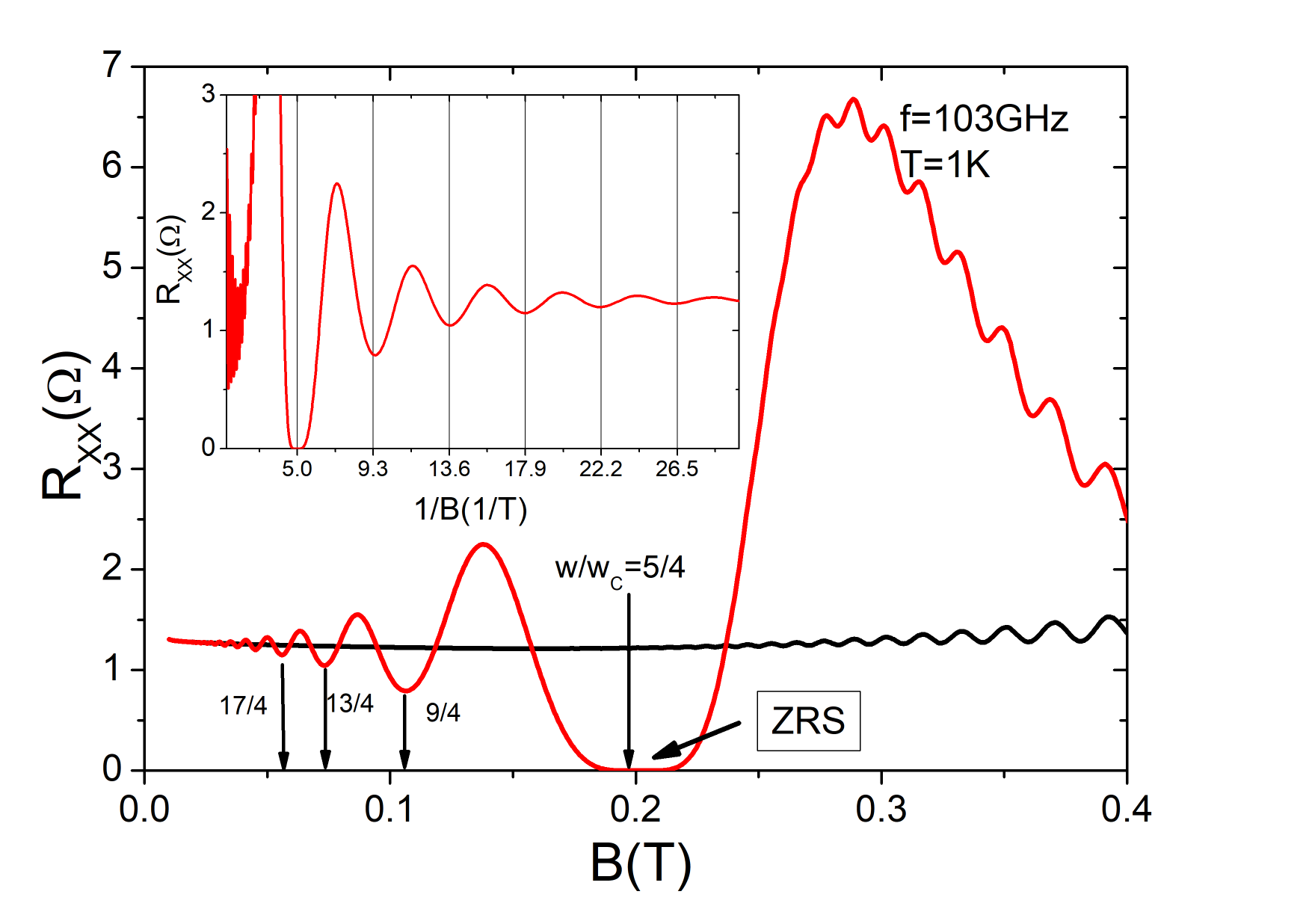}
\caption{Calculated  magnetoresistance   as a function
of $B$,  for a radiation frequency of $103$ GHz and $T=1$ K.
The dark case is also exhibited. Minima positions are indicated with arrows
corresponding to,
$\frac{w}{w_{c}}=j+\frac{1}{4}$, $j$ being a positive integer.
Zero resistance states are obtained around $B\simeq 0.2T$.
Inset: irradiated magnetoresistance showing periodicity vs $1/B$. }
\end{figure}

{\it Results}. In Fig. 4 we present calculated results of the irradiated
$R_{xx}$ vs $B$  for a radiation frequency of $103$ GHz and $T=1$ K.
The dark case is also exhibited.
In our simulations all results have been based on
experimental parameters corresponding to the experiments by Mani et al.
\cite{mani}.
 We obtain clear MIRO where the
minima positions are indicated with arrows and, as in experiments\cite{mani},
correspond to,
$\frac{w}{w_{c}}=j+\frac{1}{4}$,
$j$ being a positive integer.
Minima positions show a clear $1/4$-cycle shift, which is
a universal property that features MIRO and shows up
in any experiment about MIRO irrespective of the sort of carrier\cite{zudovhole} and
platform\cite{inamno}.  In the minima
corresponding to $j=1$, ZRS are found.
Now with the help of our present model based on coherent states we can
explain such a peculiar value for the minima position.
Thus,  it is straightforward to check out
that if we substitute   equation $\frac{w}{w_{c}}=j+\frac{1}{4}$ in
 $\Delta X_{MW}$,  (Eq. 22), we would obtain minima values
 of the latter and in turn of $R_{xx}$.
 Therefore, from the minima positions relation we can obtain
 the value $2\pi/w_{c}$ which would be
the "smoking gun" that  would reveal the presence of coherent states of
quantum harmonic oscillators  sustaining the magnetorresistance of high-quality 2DES.
Another evidence of the latter would be the MIRO periodicity with the inverse of $B$ (see inset of
Fig. 4) that would be explained by the presence of $\tau$ in the argument of the sine function.


{\it Summary}. Summing  up, we have demonstrated
that magnetoresistance in a high mobility 2DES under MW radiation can be explained in terms of the coherent states
of the quantum harmonic oscillator.
When irradiated these systems give rise to MIRO
that   reveals the presence of these quasi-classical
states in high-quality samples when under low $B$.
  These MW-driven coherent  states have
been used to calculate irradiated magnetoresistance finding that
the principle of
minimum uncertainty of coherent states is crucial to understand MIRO and
their properties   and zero resistance states.
We conclude that any experiment on
irradiated magnetoresistance of 2D systems, regardless of carrier  and
platform \cite{zudovhole,bandu},  showing MIRO reveals the existence of coherent states of the
quantum harmonic oscillator.
 We expect that dealing with even higher
mobility samples, ($\mu > 10^{7}$), it would be possible to achieve the quantum superposition
of coherent states yielding, for instance in the case of two,  even and odd coherent states
of the quantum harmonic oscillator.
Then, when irradiated,  we expect that MIRO would evolve showing striking results
revealing the presence of coherent states superposition\cite{zudov2,rui}.




This work was supported by the MCYT (Spain) grant PID2020-117787GB-I00.

\end{document}